\documentclass[12pt,letterpaper]{article}
\pdfoutput=1
\usepackage[utf8]{inputenc}
\usepackage{amsmath}
\usepackage{slashed}
\usepackage{amssymb}
\usepackage{amsthm}
\usepackage{xcolor}
\definecolor{nicered}{rgb}{0.7,0.1,0.1}
\definecolor{nicegreen}{rgb}{0.1,0.5,0.1}
\usepackage[colorlinks=true,citecolor=nicegreen,linkcolor=nicered]{hyperref}
\usepackage{graphicx}
\graphicspath{{figures/}}
\usepackage{siunitx} 
\usepackage{cancel}
\usepackage{cite}
\usepackage{soul}
\usepackage{multirow}

\definecolor{Prioridad-0}{HTML}{168F81}

\title{
Singlet-Doublet Dirac Dark Matter and Neutrino Masses
}
\author{Diego Restrepo\footnote{\href{mailto:restrepo@udea.edu.co}{restrepo@udea.edu.co}}, Andrés Rivera\footnote{\href{mailto:afelipe.rivera@udea.edu.co}{afelipe.rivera@udea.edu.co}}\\
\textit{\small  Instituto de F\'{i}sica, Universidad de Antioquia,} \\
\textit{\small  Calle 70 No. 52-21, Medell\'{i}n, Colombia}\\[4mm]
Walter Tangarife\footnote{\href{mailto:wtangarife@luc.edu}{wtangarife@luc.edu}}\\
\textit{\small Department of Physics, Loyola University Chicago,}\\
\textit{\small 1032 W. Sheridan Road, Chicago, IL, 60660, USA}
}

\date{\small \today}

\begin{document}

\maketitle
\begin{abstract}

We examine an extension of the Standard Model that addresses the dark matter puzzle and generates Dirac neutrinos masses through the radiative seesaw mechanism. 
The new field content includes a scalar field that plays an important role in setting the relic abundance of dark matter.
We analyze the phenomenology in the light of direct, indirect, and collider searches of dark matter.
In this framework, the dark matter candidate is a Dirac particle that is a mixture of new singlet-doublet fields with mass  $m_{\chi_1^0}\lesssim 1.1\,\text{TeV}$.
We find that the allowed parameter space of this model is broader than the well-known Majorana dark matter scenario.

\end{abstract}

\section{Introduction}

There is substantial evidence that supports the existence of Dark Matter (DM). Some of that evidence includes velocity dispersion in clusters of galaxies~\cite{Zwicky:1937zza} (see \cite{Gorenstein:2014iba} and for a recent review), galaxy rotation curves~\cite{Rubin:1970zza,Rubin:1980zd}, the cosmic microwave background (CMB)~\cite{Aghanim:2018eyx}, galaxy cluster collisions~\cite{clowebradacgonzalez2006}, and weak and strong gravitational lensing~\cite{Refregier:2003ct,Tyson:1998vp}. Currently, it is well established that DM makes up about $27\%$ of the energy density of the Universe, although its nature and properties remain an open puzzle.  
$N$-body simulations of early structure formation and CMB data suggest that DM is made up of cold, collisionless particles \cite{Frenk:2012ph}. 
In the light of this indication, there has been a vast exploration of candidates for DM during the last few decades, but no detection experiment has been able to find the DM particle.
In addition to the DM problem, one of the open issues in the Standard Model (SM) is the fact that neutrinos have mass, which has been confirmed by neutrino-oscillation experiments~\cite{deSalas:2017kay}. The DM problem and the neutrino mass puzzle make clear the necessity of beyond-the-standard-model physics.

In this article, we study these two puzzles within a simple extension of the Singlet-Doublet Dirac Dark Matter (SD${}^3$M) model~\cite{Yaguna:2015mva}. In Singlet-Doublet DM scenarios, a singlet and a doublet fermionic fields are added to the SM and a mixture of such fields is a Majorana DM candidate ~\cite{ArkaniHamed:2005yv,Mahbubani:2005pt,DEramo:2007anh,Enberg:2007rp,Cohen:2011ec,Cheung:2013dua,Abe:2014gua,Restrepo:2015ura,Calibbi:2015nha,Horiuchi:2016tqw,Bhattacharya:2018cgx,Bhattacharya:2018fus}. 
In the Singlet-Doublet Dirac Dark Matter model, the DM candidate is a Dirac particle, which opens a vector portal to the SM via the $Z$ gauge boson, resulting in a richer phenomenology. 
In general, this portal is not present in the Singlet-Doublet DM model with Majorana fermions, which is a generalization of the supersymmetric higgsino-bino case~\cite{Restrepo:2015ura}. 
The SD${}^3$M model addresses the DM problem while being consistent with indirect and direct experiments, as studied in Ref.~\cite{Yaguna:2015mva}. 
In addition, it can be tested in future experiments such as LZ~\cite{Mount:2017qzi} and its low mass region could be probed at the Large Hadron Collider (LHC). This simple model, however, does not generate neutrino masses. Thus, in this work, we enlarge this framework with a minimal set of scalar singlet fields in order to explain Dirac masses of SM neutrinos. These Dirac neutrino masses are generated at one-loop level in a similar fashion as in the scotogenic class of models introduced first in \cite{Ma:2006km}. An additional feature of this mechanism is the enhancement of the scalar portal that is suppressed in the minimal framework of the SD$^3$M model studied in Ref.~\cite{Yaguna:2015mva}.

We describe our model in Sec.~\ref{sec:model}. In Sec.~\ref{sec:neutrinos}, we present the generation of the neutrino masses. Sec.~\ref{sec:DM} includes the DM analysis and numerical results, and we close with Conclusions.

\section{Description of the model}
\label{sec:model}
In this model, we extend the symmetry of the Standard Model (SM) with two discrete symmetries, $\mathbb{Z}_2$ and $\mathbb{Z}_2^{'}$. $\mathbb{Z}_2$ stabilizes the DM particle and $\mathbb{Z}_2^{'}$ forbids the generation of neutrino masses via the seesaw mechanism at tree level~\cite{Chulia:2016ngi,CentellesChulia:2018gwr}. All SM particles are even under these discrete symmetries. This model also includes the following additional fields: A real scalar singlet $S=(S^0+v_S)/\sqrt{2}$, two real scalar singlets $\sigma_i$ (which are needed to obtain a rank-2 neutrino mass matrix), two chiral fermionic singlets $\psi_{L}$ and $\psi_{R}$, one Dirac SU(2) vector-like fermion $\Psi$ with hypercharge $-1/2$, and three right-handed neutrinos $\nu_R^\alpha$. In addition, we assume that global $U(1)_{B-L}$ is conserved and that the new fermions are charged under this symmetry. A result of this assumption is that Majorana mass terms are forbidden, leading to Dirac neutrino masses. The particle content is also listed in Table~\ref{tab:particle-content-Dirac}~\footnote{A different $U(1)_{B-L}$ charge assignment, in radiative Dirac neutrino mass models, was made in Ref.~\cite{Calle:2018ovc} for the case of complex $\sigma_i$.}.

\begin{table}
\centering
\begin{tabular}{|c|c|c|c|c|}
\hline
Leptons and scalars fields & $(\text{SU(2)}_L, \text{U(1)}_Y)$ & $\mathbb{Z}_2$ (DM) 
& $\mathbb{Z}_2^{'}$ & $U(1)_{B-L}$\\
\hline
$L_\beta=\begin{pmatrix}
 \nu_L \\ l_L
\end{pmatrix}_{\beta}$ &$(2,-1/2)$&+&+ &-1\\
$l_R^\alpha$ &$(1,0)$&+&+& -1\\
$H=\bigg( H^+, \dfrac{h^0+v}{\sqrt{2}}\bigg)^T$ & $(2,1/2)$ & + & + & 0\\
\hline
$S$ & $(1,0)$ & + & - & 0\\
$\sigma_i$ & $(1,0)$ & - & - & 0\\
\hline
$\psi_{L}$ & $(1,0)$ & - & + & -1\\
$\psi_{R}$ & $(1,0)$ & - & - & -1\\
$\Psi=\begin{pmatrix}
\Psi^0 \\ \Psi^-
\end{pmatrix}$ & $(2,-1/2)$ & - & - & -1\\
$\nu_{R }^\alpha$ & $(1,0)$ & + & - & -1\\
 \hline
\end{tabular}
\caption{Particle content of the model. }
\label{tab:particle-content-Dirac}
\end{table}
The most general Lagrangian, invariant under the symmetries mentioned above, contains the terms
\begin{align}
\label{eq:full-lagrangian-Dirac}
\mathcal{L} \supset &
-M_{\Psi}\, \overline{\Psi}\Psi -V(H,\sigma_i, S)\nonumber  \\
&+  \left[ h_a^{\beta i} \overline{L}_\beta \Psi \sigma_i 
+ h_b^{\alpha i}\, \overline{{\psi}_{L}} \nu_{R\alpha} \sigma_i
+ h_c\, \overline{{\psi}_{R}}\psi_{L} S 
+ h_d\,\overline{\Psi}\widetilde{H}\psi_{R} 
+ \text{h.c.}\right] \,, 
\end{align}
where $h$'s are Yukawa couplings, which we assume to be real parameters for the sake of simplicity, and $\widetilde{H}=i\sigma_2 H^*$. 
Notice that the vector-like fermion $\Psi$ can be written in terms of two chiral doublets $\Psi_L=(\Psi_L^0,\Psi_L^-)^T$ and $\widetilde{(\Psi_R)}=(-(\Psi_R^-)^{\dagger},(\Psi_R^0)^{\dagger})^T$ with opposite hypercharge~\cite{Calle:2018ovc}, as shown in Appendix~\ref{sec:l-weyl-spinors}. 

The scalar potential is given by
\begin{align}
\label{eq:ScalarPotential}
V(H,\sigma_i,S) =& -\mu^2 H^{\dagger}H + \dfrac{\lambda_1}{2}(H^{\dagger}H)^2 + \dfrac{1}{2}m_{\sigma_i}^2 \sigma_i^2 + \lambda^{\sigma H}_i H^{\dagger}H \sigma_i^2
+ \dfrac{\lambda_i^{\sigma}}{2} \sigma_i^4 \nonumber \\
&+ \dfrac{1}{2} m_S^2 S^2 + \lambda_{SH} H^{\dagger}H S^2 
+ \lambda^{S\sigma_i}S^2\sigma_i^2
+ \dfrac{\lambda^S}{2} S^4 \,.
\end{align}
The condition that the potential is bounded from below is fulfilled by imposing $\mu^2 >0$, $m_{\sigma i}^2 >0$, $m_S^2 >0$, together with the co-positivity of the potential~\cite{Kannike:2012pe}, which yields
\begin{align}
\label{eq:cond-bounded}
&\lambda_1 \geq 0\,,\hspace{0.4 cm} \lambda_i^{\sigma} \geq 0\,,
\hspace{0.4 cm} \lambda^S \geq 0\,,\\
&\dfrac{\lambda_i^{\sigma H}}{2}+\sqrt{\lambda_1\lambda_i^\sigma}  \geq 0\,,\hspace{0.4 cm} 
\dfrac{\lambda_{SH}}{2}+\sqrt{\lambda_1\lambda^S} \geq 0\,,\hspace{0.4 cm}
\dfrac{\lambda^{S\sigma}}{2}+\sqrt{\lambda_i^{\sigma}\lambda^S} \geq 0\,,\\
&\sqrt{\lambda_1\lambda_i^\sigma \lambda^S} +\dfrac{\lambda_i^{\sigma H}}{2}\sqrt{\lambda^S}
 +\dfrac{\lambda_{SH}}{2}\sqrt{\lambda_i^\sigma}
 +\dfrac{\lambda_i^{S\sigma_i}}{2}\sqrt{\lambda_1}+\nonumber\\
 &\sqrt{2\bigg(\dfrac{\lambda_i^{\sigma H}}{2}+\sqrt{\lambda_1\lambda_i^\sigma}\bigg)
 \bigg(\dfrac{\lambda_{SH}}{2}+\sqrt{\lambda_1\lambda^S}\bigg)
 \bigg(\dfrac{\lambda^{S\sigma_i}}{2}+\sqrt{\lambda_i^\sigma\lambda^S}\bigg)}\geq 0\,.
\end{align}
These conditions are trivially satisfied if we demand that all $\lambda$'s are positive.

\subsection{Symmetry breaking and spectrum}

The scalar potential \eqref{eq:ScalarPotential} allows a vacuum expectation value (VEV) for the singlet scalar, $\langle S \rangle = v_S/\sqrt{2}$, in addition to the Higgs VEV, $\langle H \rangle =v/\sqrt{2}$. These VEVs are given by the tadpole equations
\begin{align}
\label{eq:tadpoles}
t_{H} &= \left(\dfrac{\partial V}{\partial v}\right)
= -\mu ^2 v +\frac{\lambda _1 v^3}{2} + \dfrac{\lambda_{SH}}{2} v v_S^2=0\,, \\
t_S &= \left(\dfrac{\partial V}{\partial v_S}\right)
= m_S^2 v_S + \lambda_{SH}v^2 v_S+ \lambda^S v_S^3=0\,,
\end{align}
which are used to eliminate the parameters $\mu$ and $m_S$. The scalar spectrum contains the $\mathbb{Z}_2$-even scalars $h^0$, $S^0$, and  $\mathbb{Z}_2$-odd scalars $\sigma_i$. In the basis $(h^0,S^0)$, the mass matrix for the $\mathbb{Z}_2$-even scalars is given by
\begin{equation}
\label{eq:mh-matrix}
m_h^2=
\left(
\begin{array}{cc}
 -\mu ^2+\frac{1}{2} v_S^2 \lambda_{SH}+\frac{3 \lambda _1 v^2}{2} & v v_S
   \lambda_{SH} \\
 v v_S \lambda_{SH} & \frac{m_s^2}{2}+\frac{3}{2} \lambda ^S v_S^2+\frac{v^2
   \lambda_{SH}}{2} \\
\end{array}
\right)\,,
\end{equation}
which is diagonalized by a unitary transformation
\begin{equation}
\label{eq:ZH-matrix}
Z^H m_h^2Z^{H\dagger} = m_{h,\text{diag}}^{2}\,,
\end{equation}
such that
\begin{equation}
\begin{pmatrix}
h_0 \\ S^0
\end{pmatrix} =
Z^H
\begin{pmatrix}
h_1 \\ h_2
\end{pmatrix}
 =
\begin{pmatrix}
\cos\alpha & \sin\alpha \\
-\sin\alpha & \cos\alpha 
\end{pmatrix}
\begin{pmatrix}
h_1 \\ h_2
\end{pmatrix}\,. \label{eq:Higgs-mixing}
\end{equation}
The lightest eigenstate, $h_1$, is identified with the SM Higgs boson, whereas the heavier one will be a heavy Higgs boson not yet discovered at the LHC. 
The existence of a second Higgs can be beneficial in order to stabilize the metastable electroweak vacuum of the SM, as argued in Ref.~\cite{Falkowski:2015iwa}. However, some constraints need to be taken into account. The Higgs-boson mixing \eqref{eq:Higgs-mixing} generates the effective interaction terms
\begin{equation}
\label{eq:higgs-couplings}
\mathcal{L} \supset \dfrac{h_1\cos\alpha+h_2\sin\alpha}{v}
\left(2m_W^2W_{\mu}^{+}W^{\mu -}+m_Z^2Z_{\mu}Z^{\mu}-\sum_f m_f \bar{f}f\right)\,,
\end{equation}
which suppress the partial decay of $h_1$ to SM fields by factor $\sim\cos^2\alpha$.
Similarly, the heavier scalar $h_2$ could have a decay width $\Gamma(h_2\to h_1h_1)\sim \sin^2\alpha$ if it is kinematically allowed. 
In addition, $h_2$ is constrained by the electroweak oblique parameters since, for $m_{h_2} \gg m_{h_1}$, it has been shown that~\cite{Falkowski:2015iwa}
\begin{align}
\label{eq:T}
T& \approx-\dfrac{3}{8\pi\cos^2\theta_W}\sin^2\alpha\log(m_{h_2}/211\,\text{GeV})\,,\\
S& \approx\dfrac{1}{6\pi}\sin^2\alpha\log(m_{h_2}/81\,\text{GeV})\,.
\end{align}
Further constraints are provided by LEP and LHC searches for Higgs-like scalars. For instance, processes such as $h_i\to\gamma\gamma$, $h_2\to ZZ$, $h_2\to WW$, etc, have been analyzed in the literature~\cite{Aaboud:2018xdt,ATLAS:2018uso,Sirunyan:2018ouh,Khachatryan:2015cwa}.
As shown in Ref.~\cite{Falkowski:2015iwa}, by combining the experimental constraints and taking care of the vacuum stability in the evolution of the renormalization group equations up to the Planck scale, these observables and constraints are under control if we demand a mixing $|\sin\alpha| \lesssim 0.3$, which has been taken into account in this work.
On the other hand, the $\mathbb{Z}_2$-odd scalar sector is assumed to be already in the diagonal basis,
\begin{equation}
\label{eq:ms-matrix}
m_\sigma^2=
\left(
\begin{array}{cc}
 m_{\text{$\sigma $1}}^2+ v^2 \lambda_1^{\sigma_1 H} 
 + v_S^2 \lambda^{S\sigma_1}& 0 \\
 0 & m_{\text{$\sigma $2}}^2+ v^2 \lambda _2^{\sigma_2 H}
 + v_S^2 \lambda^{S\sigma_2} \\
\end{array}
\right)\,.
\end{equation}
While the lightest of these scalars could be a suitable candidate for DM, in this work we focus instead on fermionic DM. The scalar DM phenomenology is expected to be rather similar to the one in the Majorana version for both DM and neutrino masses~\cite{Restrepo:2013aga, Restrepo:2015ura}. Therefore, we will assume the $\sigma_i$ fields to be heavy ($m_{\sigma_i} > 1$ TeV) while playing  an important role only in the generation of neutrino masses, as shown in the next section.

Regarding the $\mathbb{Z}_2$-odd fermionic sector, this model contains one charged Dirac fermion $\Psi^{\pm}$ with mass $M_{\Psi}$ and two neutral Dirac fermions, $\chi_j^0\, (j=1,2)$.  
In the basis $N_{Li}=(\Psi_{L}^0,\psi_{L})$, $N_{Ri}^{\dagger}=((\Psi_R^0)^{\dagger},(\psi_{R})^{\dagger}),$ the fermionic mass matrix is given by
\begin{equation}
\label{eq:mchi-matrix}
m_{\psi^0}=
\left(
\begin{array}{cc}
 M_{\Psi } & \frac{h_d v}{\sqrt{2}} \\
 0 & M_N \\
\end{array}
\right)\,,
\end{equation}
where $M_N=h_c v_S/\sqrt{2}$ is the Dirac mass term for $\psi_{L,\,R}$, which results after the $\mathbb{Z}_2^{'}$ symmetry breaking.  
This matrix is diagonalized by the bi-unitary transformation
\begin{equation}
\label{eq:mchi-diag}
V^*m_{\psi^0}U^{\dagger} = m_{\chi_i^0}^{\text{diag}}\,,
\end{equation}
where the mass eigenstates, $\chi_j^0=(\chi_{L},\chi_{R}^{\dagger})_j$,  are defined by
\begin{equation}
\label{eq:chi-fermions}
\chi_{Lj} 
= V_{ji}N_{Li} 
= \begin{pmatrix}
 \cos\theta_L & \sin\theta_L \\
 -\sin\theta_L & \cos\theta_L
\end{pmatrix}
\begin{pmatrix}
\Psi_{L}^0 \\ \psi_{L}
\end{pmatrix},
\hspace{0.5 cm} 
\chi_{Rj}^{\dagger}
= U_{ji}N_{Ri}^{\dagger}
 =\begin{pmatrix}
 \cos\theta_R & \sin\theta_R \\
 -\sin\theta_R & \cos\theta_R
\end{pmatrix}
\begin{pmatrix}
(\Psi_R^0)^{\dagger} \\ (\psi_{R})^{\dagger}
\end{pmatrix} \,,
\end{equation}
where $\theta_{L,R}$ are mixing angles.  
In this work, the lightest of these Dirac fermions, $\chi_1^0$, is the candidate for the DM particle. 
Notice our choice to parametrize the fermionic sector using $m_{\chi_1^0}$, $m_{\chi_2^0}$, $\theta_L$, and $\theta_R$, instead of $M_\Psi$, $h_c$, $h_d$, and $v_S$.

\section{Dirac neutrino masses}
\label{sec:neutrinos}

\begin{figure}[t]
\begin{center}
\includegraphics[scale=0.5]{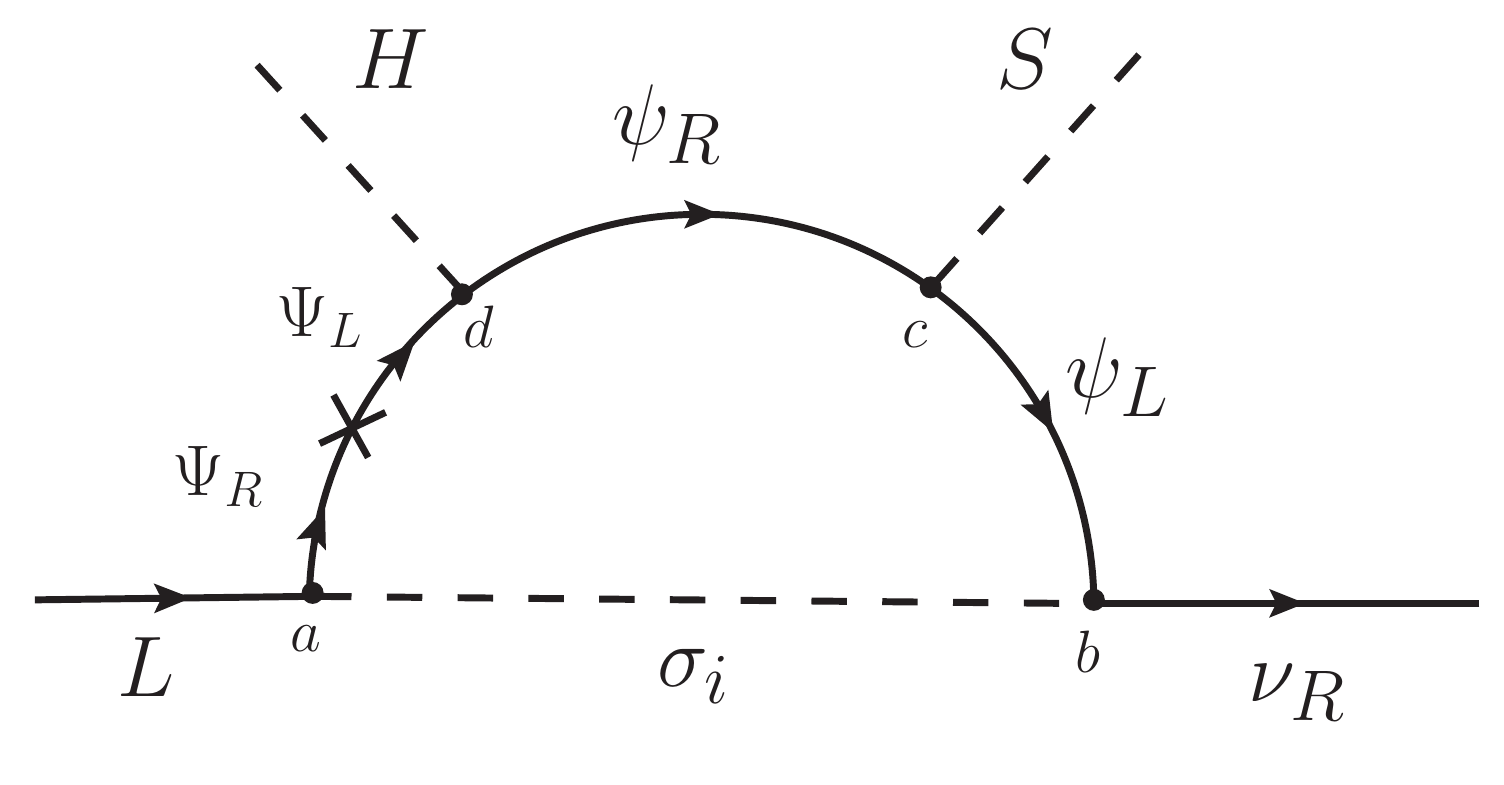}
\end{center}
\caption{One-loop generation of Dirac neutrino masses. 
The arrows represent the flow of $U(1)_{B-L}$ charges.}
\label{fig:neutrino-GB}
\end{figure}

In this framework, the scalars $H$ and $S$ acquire VEVs. As a result of this symmetry breaking, neutrinos get masses via the five-dimensional effective operator
\begin{equation}
\label{eq:L5-operator}
\mathcal{L}_5^{D}= -\dfrac{g_{\alpha\beta}}{\Lambda}\bar{L}_{\alpha}\widetilde{H}\nu_{R\beta}S + \text{h.c.}\,,
\end{equation}
which is generated at the one-loop level. Ref.~\cite{Yao:2018ekp} has performed a systematic study of the one-loop topologies that give rise to this operator\footnote{In particular, the model proposed in this work is similar to the topology T1-2-A-I ($\alpha=0$) in Ref.~\cite{Yao:2018ekp}. However, in that case all new fermions are vector-like. Instead, we use chiral fermions with fewer degrees of freedom.}. In our specific scenario, Dirac neutrino masses arise from the one-loop diagram shown in Fig.~\ref{fig:neutrino-GB}.
In the limit of low neutrino momentum, that diagram yields the mass matrix
\begin{align}
\label{eq:neutrino-mij-matrix}
\mathcal{M}_{\alpha\beta}=& \sum_{i=1}^2\sum_{j=1}^2
\dfrac{U_{j1}V_{j2}}{16\pi^2}
\times h_b^{\alpha i} h_a^{\beta i} m_{\chi_j^0} 
\times \left[\dfrac{m_{\chi_j^0}^2 \ln(m_{\chi_j^0}^2) - m_{\sigma_i}^2 \ln(m_{\sigma_i}^2 )}{\left(m_{\chi_j^0}^2-m_{\sigma_i}^2\right)}\right]\nonumber , \\
=&\sum_{i=1}^2
h_b^{\alpha i}\times \Lambda_i\times h_a^{\beta i}\,,
\end{align}
where $\Lambda_i$ is the loop factor, defined as
\begin{align}
\label{eq:Lambda}
\Lambda_i = & \sum_{j=1}^2
\dfrac{U_{j1}V_{j2}}{16\pi^2}
\times  m_{\chi_j^0} 
\times \left[\dfrac{m_{\chi_j^0}^2 \ln(m_{\chi_j^0}^2) - m_{\sigma_i}^2 \ln(m_{\sigma_i}^2 )}{\left(m_{\chi_j^0}^2-m_{\sigma_i}^2\right)}\right]\nonumber ,\\
= & \sum_{j=1}^2
\dfrac{U_{j1}V_{j2}}{16\pi^2}
\times  m_{\chi_j^0} 
\times \left[\dfrac{m_{\chi_j^0}^2}{\left(m_{\chi_j^0}^2-m_{\sigma_i}^2\right)}\ln{\left(\dfrac{m_{\chi_j^0}^2}{m_{\sigma_i}^2}\right)}\right]\,.
\end{align}
In the last equation, we used the relation 
\begin{equation}
\label{eq:sumcero}
\sum_{j}^2 m_{\chi_j^0} U_{j1}V_{j2}  = 0\,,
\end{equation}
which is a consequence of Eqs.~\eqref{eq:mchi-matrix} and ~\eqref{eq:mchi-diag}. 

We need to set the correct Yukawa couplings in the Lagrangian~\eqref{eq:full-lagrangian-Dirac} in order to reproduce the current neutrino oscillation data to $3\sigma$~\cite{deSalas:2017kay}. That is, we need to invert the problem and use the neutrino parameters to choose our Yukawa couplings. This can be done by using the fact that, in the basis where $\nu_R^{\alpha}$ are mass eigenstates, the neutrino mass matrix can be written as~\cite{Kanemura:2011jj}
\begin{equation}
\label{eq:PMNS-relation}
\mathcal{M}_{\alpha\beta}=(U_{\text{PMNS}})_{\alpha\beta}\,(m_{\nu})_{\beta}\,,
\end{equation}
where $U_{\text{PMNS}}$ is the Pontecorvo-Maki-Nakagawa-Sakata matrix~\cite{Maki:1962mu} and $m_{\nu}$ are the neutrino mass eigenvalues.
It is well known that current neutrino oscillation data allow for normal or inverted ordering, $m_{\nu 1}<m_{\nu 2}<m_{\nu 3}$ or $m_{\nu 3}<m_{\nu 1}<m_{\nu 2}$, respectively.
In this work, we choose the normal ordering. Using the Eqs.~\eqref{eq:neutrino-mij-matrix} and~\eqref{eq:PMNS-relation}, we obtain twelve unknown parameters, $h_a^{\beta i}, h_b^{\alpha i}$, with nine equations. We can further simplify our analysis by imposing $m_{\nu 1}=0$, which allows us to set $h_a^{1i}=0$, leaving the couplings $h_a^{2i}$ and $h_a^{3i}$ as free parameters.
With these assumptions, we obtain the following relations:
\begin{eqnarray}
\label{eq:ha-and-hb}
h_a^{1i}&=&0\,, \nonumber\\
h_a^{2i,3i}&=& \text{free}\,, \nonumber\\
h_b^{\alpha 1}&=& -\dfrac{1}{\Lambda_1}
\left(\dfrac{h_a^{32} m_{\nu 2} U_{\alpha 2} - h_a^{22} m_{\nu 3} U_{\alpha 3}}{h_a^{22} h_a^{31}- h_a^{21} h_a^{32} }\right), \nonumber\\
h_b^{\alpha 2}&=& -\dfrac{1}{\Lambda_2}
\left(\dfrac{h_a^{31} m_{\nu 2} U_{\alpha 2} - h_a^{21} m_{\nu 3} U_{\alpha 3}}{h_a^{22} h_a^{31}- h_a^{21} h_a^{32} }\right)\,.
\end{eqnarray}
It is noteworthy that, with this choice of parameters, some lepton-flavor-violation (LFV) processes such as $\mu\to e\gamma$ are suppressed since they are proportional to the $h_a^{1i}$ coupling. However, other processes, like $\tau\to \mu\gamma$, are still allowed with much lower experimentally restrictions.

\section{Dark matter}
\label{sec:DM}

In this work, the Dirac fermion $\chi_1^0$ is the DM candidate while the scalars $\sigma_i$ are chosen to be much heavier than $\chi_j^0$. In this section, we discuss the main process that sets the relic abundance of DM as well as the direct detection of such a particle. 

\subsection{Dark matter relic density}
\label{sec:h2-relic-density}

\begin{figure}
\centering
\includegraphics[scale=0.5]{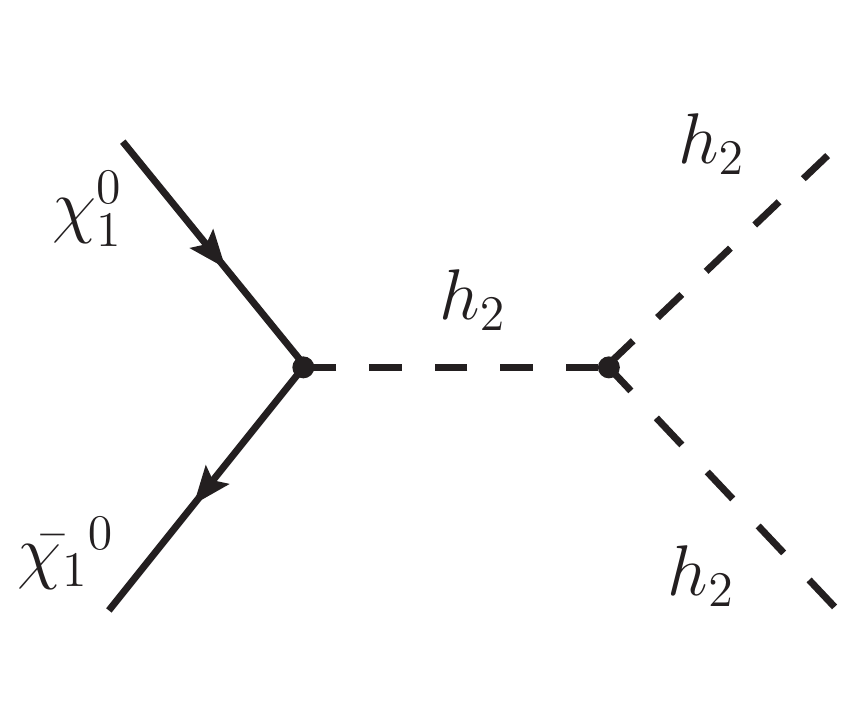} \hspace{.5 cm}
\includegraphics[scale=0.5]{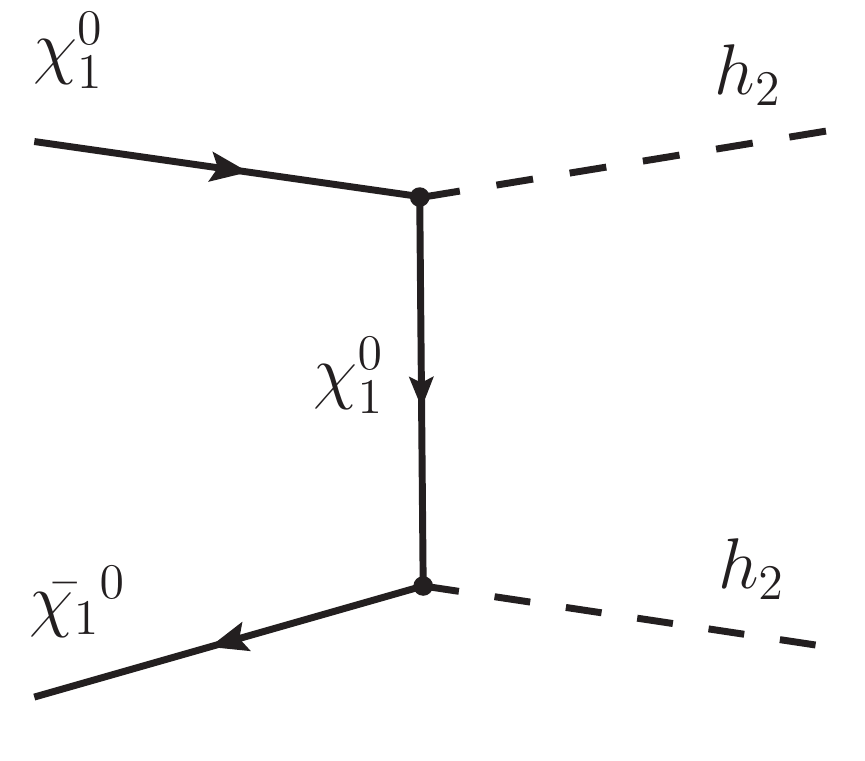} \hspace{.5 cm}
\includegraphics[scale=0.5]{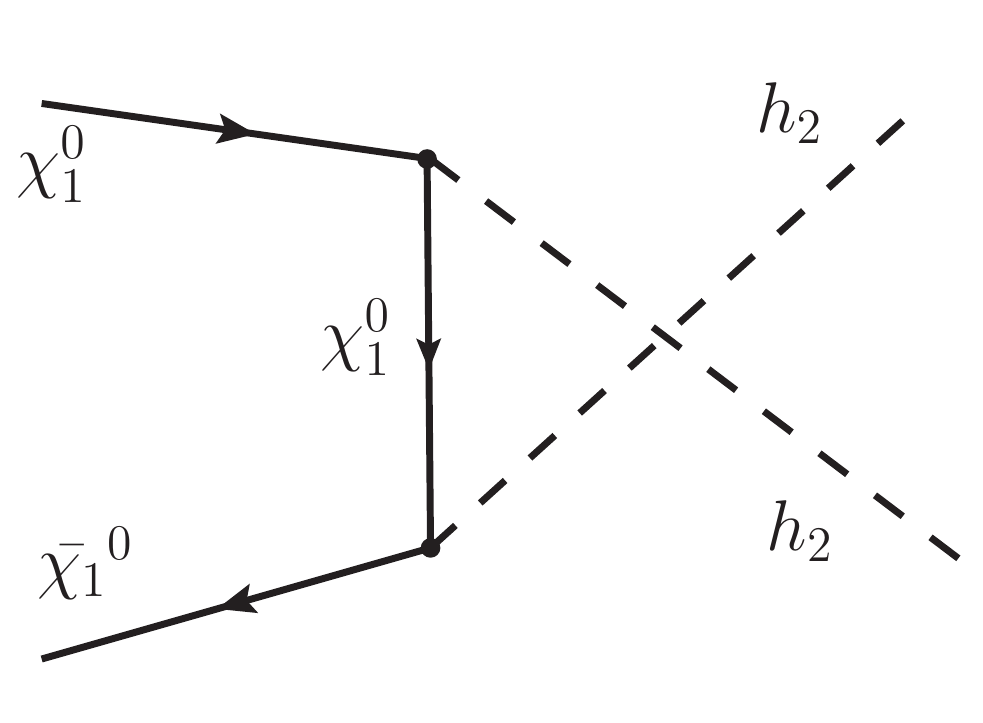}

\caption{Diagrams that contribute to the annihilation process $\chi_1^0\bar{\chi}_1^0\to h_2h_2$.}
\label{fig:DM-Annihilation}
\end{figure}

In the class of models that we study in this article, $\chi_1^0$ couples to the Higgs and to the $Z$ boson through the singlet-doublet mixing. This implies that the couplings of the DM particle to the $Z$ vector are largely constrained by direct detection experiments, leading to a mostly singlet DM candidate as seen numerically in the next section. In Ref.~\cite{Yaguna:2015mva}, this fact restricted the allowed parameter space to quasi-degenerate mass eigenstates for the fermionic fields and the DM abundance was determined mainly through coannihilations. In our work, the presence of the additional scalar $S$ adds new annihilation channels, opening up the range of masses for the fermions and providing a richer phenomenology. Specifically, the processes involved in the calculation of the DM relic abundance include $\chi_i^0\bar{\chi_j}^0\to h_kh_l$,  
$\chi_i^0\bar{\chi_j}^0\to W^+W^-$,  
$\chi_i^0\bar{\chi_j}^0\to ZZ$,
$\chi^{\pm}\chi^{\mp}\to f\bar{f}$,
$\chi_i^0\chi^{+}\to f\bar{f'}$,
$\chi_i^0\chi^{\pm}\to A/Z W^{\pm}$, and
$\chi^{\pm}\chi^{\mp}\to W^+W^-$.  As explained in the next section, our numerical analysis takes into account all these channels; however, the most relevant process is $\chi_1^0\bar{\chi}_1^0\to h_2h_2$, which gets contributions from the diagrams shown in Fig.~\ref{fig:DM-Annihilation}.

The early thermal evolution of our DM candidate follows the standard WIMP freeze-out mechanism. In the initial state, the DM species was in thermal equilibrium with the rest of particles in the universe. As the universe adiabatically cools down to a temperature below the DM mass, the DM annihilation rate is overtaken by the expansion of the universe, $\Gamma \ll H$, and a relic density of DM is frozen-out. The current relic abundance of DM is computed by solving the Boltzmann equation, which yields~\cite{Kolb:1990vq}
\begin{equation}
\Omega_\chi \,=\,2\sqrt{\frac{4 \pi^{3} g_{*}(m_\chi)}{45}} \frac{8 \pi}{90 H_{0}^{2}} \frac{x_{f}}{\langle\sigma v\rangle} \frac{T_{0}^{3}}{M_{\mathrm{Pl}}^{3}}\,,
\end{equation} where $\langle\sigma v\rangle$ is the thermally-averaged annihilation cross-section, $g_*(m)$ is the effective number of degrees of freedom at $T=m$, and $x_f\equiv m/T_{\rm freeze-out}$. The factor of $2$ in front of the right-hand side of the equation above is due to the fact that we have a Dirac particle and $n_{\rm DM}= n_{\chi}+n_{\bar{\chi}}$~\cite{Srednicki:1988ce}. The partial-wave expansion of the annihilation cross-section, $\langle \sigma v\rangle \, \approx\, a\,+\,b v^2\,+\,\mathcal{O}(v^4)$, leads to the well-known expression
\begin{equation}
\label{eq:RelicAbundance}
\Omega_{\chi} h^{2} \approx 2\frac{1.04 \times 10^{9} x_{f}}{M_{\mathrm{Pl}} \sqrt{g_{*}(m_\chi)}\left(a+3 b / x_{f}\right)}\,,
\end{equation}
where $h$ is today's Hubble parameter in units of $100~{\rm km/s/Mpc}$. The $\chi_1^0\bar{\chi}_1^0\to h_2h_2$ annihilation cross-section has no $s-$wave contribution, which means that $a=0$. In order to achieve the measured relic density, $\Omega_{\chi} h^{2} =0.1200 \pm 0.0012$~\cite{Aghanim:2018eyx}, the annihilation cross-section is required to be approximately $\langle \sigma v \rangle\sim 3\times 10^{-26}$ cm$^{3}$s$^{-1}$. For illustrative purposes, let us write the specific expression for the cross-section in the limit where DM is purely singlet, $\chi^0_j = \left( \begin{array}{l}{\psi_L} \\ {\psi_R^\dagger}\end{array}\right) $ :
\begin{eqnarray}
\langle\sigma v\rangle &=& \frac{ h_c^2\sqrt{1-\mu^2}}{16 \pi m_{\chi_0}^2}\left( \frac{9 (\lambda_{SH})^2 v_S^2}{16 m_{\chi_0}^2(\mu^2-4)^2}-\frac{ h_c \lambda_{SH} v_S (20-13\mu^2+2\mu^4)}{2\sqrt{2} m_{\chi_0}(\mu^2-4)^2(\mu^2-2)^2}  +\frac{h_c^2 (9-8\mu^2+2\mu^4)}{6 (\mu^2-2)^2}\right) v^2, \nonumber \\
 &=& b \,v^2,
\end{eqnarray}
where $\mu \equiv m_{\chi_0} / m_{h_2} < 1$. The first term corresponds to the $s-$channel while the last one comes from the $t$ and $u$ channels and their interference. The second term results from the interference between the $s$ and the $t,\,u$ channels (see Fig. \ref{fig:DM-Annihilation}). In the next section, we present the numerical results of this computation and the corresponding relic abundance. Finally, let us mention that there is a clear consequence of having $p-$wave annihilation of DM for indirect-detection searches. Since $\sigma v \propto v^2$, the annihilation rate is suppressed by several orders of magnitude in the low-velocity limit (today) compared to the value in the early universe, escaping the bounds from current indirect searches, which require $\sigma v \lesssim 3\times 10^{-26}$ cm$^{3}$s$^{-1}$.

\subsection{Direct detection of DM}
\label{{sec:DD}}

\begin{figure}
\centering
\includegraphics[scale=0.5]{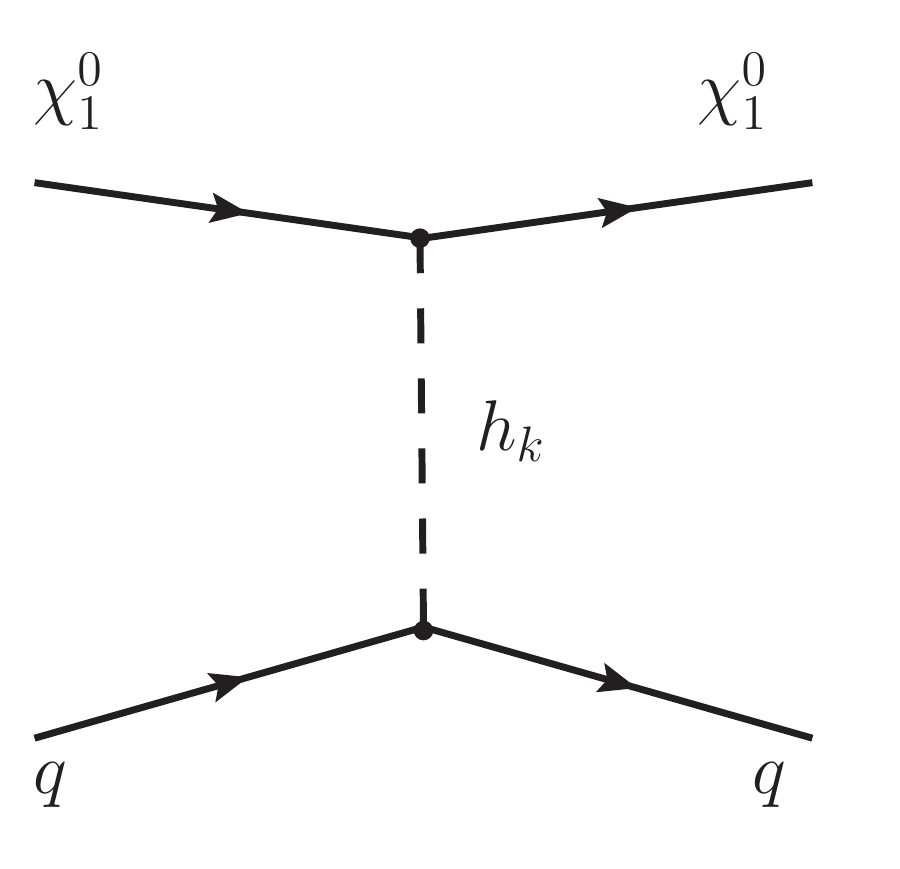} \hspace{2 cm}
\includegraphics[scale=0.5]{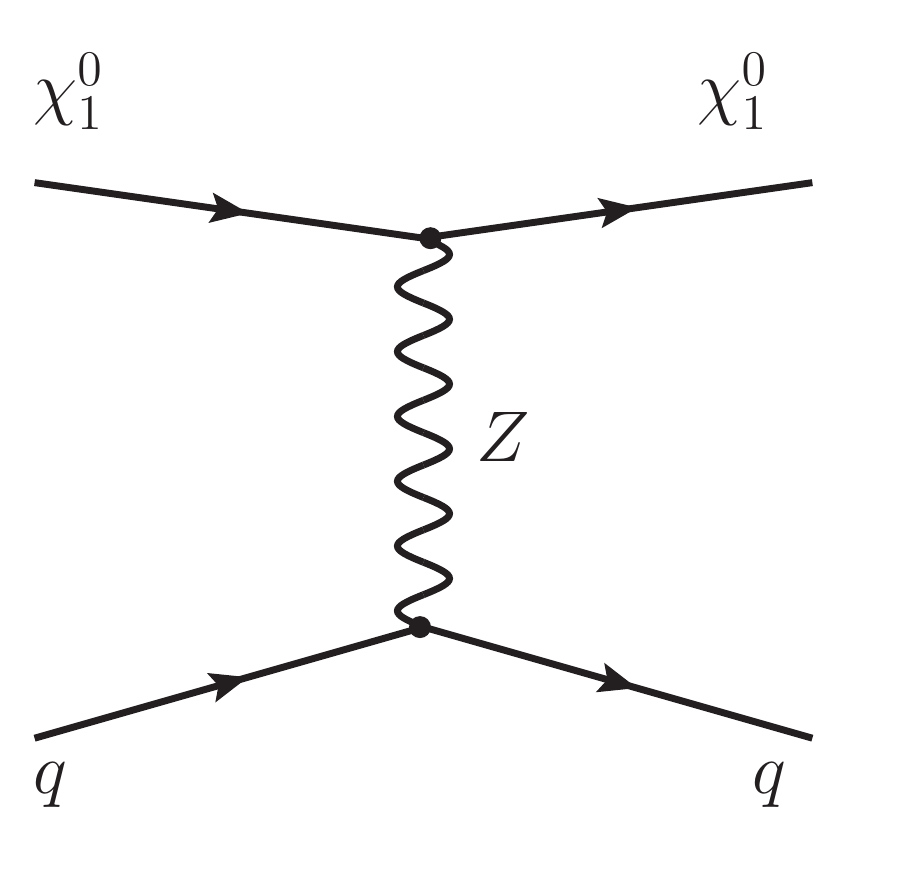}
\caption{SI independent DM-nucleon interactions: Scalar (left) and vector (right) portals.}
\label{fig:SI-diagrams}
\end{figure}

As mentioned above, since $\chi_1^0$ couples to scalars and the $Z$ boson, there are direct and indirect detection restrictions that can be imposed on this model. Regarding elastic scattering of $\chi_1^0$ with nuclei, we have two different contributions, the scalar/vector or spin-independent (SI) interaction and the axial-vector or spin-dependent (SD) interaction. It is noteworthy that in the SUSY analog of the Singlet-Doublet model, i.e. the higgsino-bino model, the SI interaction is only due to the scalar portal. In that case, the vector portal with the $Z$ boson is closed since the DM particles are Majorana fermions. 
However, in our scenario, the SI interaction of DM with nucleons contains both portals: a $t$-channel mediated by the Higgses $h_k$ and a $t$-channel mediated by the $Z$ gauge boson, which correspond to the diagrams shown in Fig.~\ref{fig:SI-diagrams}. We use the standard nucleon-form-factor formalism to incorporate these processes into the WIMP-nucleon amplitudes~\cite{Belanger:2008sj}. 
Given the interaction Lagrangian $\mathcal{L}^{SI}_{e,o}= \lambda_{N,e}\bar{\psi}_\chi\psi_\chi\psi_N\psi_N + \lambda_{N,o}\bar{\psi}_\chi\gamma^\mu\psi_\chi\psi_N\gamma_\mu\psi_N $,  $N=p,n$, the scattering cross-section per nucleus is given by
\begin{equation}
\label{eq:sigma0}
\sigma_0^{SI} = \dfrac{4\mu_\chi^2}{\pi}\left(\lambda_pZ+\lambda_n(A-Z)\right)^2\,,
\end{equation}
where $\mu_\chi=m_{\chi}M_A/(m_{\chi}+M_A)$ is the WIMP-nucleus reduced mass, $Z$ is the nucleus charge, $A$ is the total number of nucleons, and $\lambda_p , \lambda_n$ are related to $\lambda_{N,e}, \lambda_{N,o}$ as we will show in the next paragraph. When implementing experimental bounds, the relevant quantity is the scattering cross-section per nucleon, which is written as
\begin{align}
\label{eq:sigmaN}
\sigma_N^{SI} = \dfrac{m_N^2}{\mu_\chi^2 A^2}\sigma_0^{SI}\,,
\end{align}
where $m_N$ is the nucleon mass.

Using the nucleon-quark operator formalism, $\lambda_p , \lambda_n$ are found to be~\cite{Belanger:2008sj}
\begin{align}
\label{eq:lp-ln}
\lambda_p &= \dfrac{\lambda_{p,e}\pm\lambda_{p,o}}{2}
=\left( g_{\chi_1^0\chi_1^0 h_k}\dfrac{1}{m_{h_k}^2}\dfrac{m_p\sum_{q}f_q^p}{v} 
\pm\sum_{q=u,d}f^p_{Vq}\lambda_{q,o}\right)/2 \,, \\
\lambda_n &= \dfrac{\lambda_{n,e}\pm\lambda_{n,o}}{2}
=\left( g_{\chi_1^0\chi_1^0 h_k}\dfrac{1}{m_{h_k}^2}\dfrac{m_n\sum_{q}f_q^n}{v} 
\pm\sum_{q=u,d}f^n_{Vq}\lambda_{q,o}\right)/2\,,
\end{align}
where the $+(-)$ signs correspond to WIMP (anti-WIMP) interaction, $\sum_{q}f_q^p \approx \sum_{q}f_q^n = f_N \approx 0.3$ is the form factor for the scalar interaction~\cite{Alarcon:2011zs,Alarcon:2012nr}, $f_{Vq}^N$ counts the number of quarks $u,d$ inside the nucleon ($f_{Vu}^p=2, f_{Vd}^p=1$, $f_{Vu}^n=1, f_{Vd}^n=2$) and $\lambda_{q,o}$ are the vector form factors, which, in our model, follow the relations
\begin{align}
\label{eq:fV-factors}
\sum_{q=u,d}f^p_{Vq}\lambda_{q,o}&=\dfrac{M_Z (\cos^2\theta_L+\cos\theta_R^2)}{2\,v}\times \dfrac{1}{M_Z^2}\times \dfrac{e}{4\sin\theta_W\cos\theta_W}(1-4\sin\theta_W^2) \,,\\
\sum_{q=u,d}f^n_{Vq}\lambda_{q,o}&=\dfrac{M_Z (\cos^2\theta_L+\cos\theta_R^2)}{2\,v}\times \dfrac{1}{M_Z^2}\times \dfrac{(-e)}{4\sin\theta_W\cos\theta_W}\,.
\end{align}
In the above formulas, $\theta_W$ is the weak-mixing angle and $\theta_{L,R}$ are the mixing angles defined in Eq.~\eqref{eq:chi-fermions}.  
Therefore, the total SI cross-section can be written as
\begin{align}
\label{eq:full-SI}
\sigma_N^{SI} = \sigma^{SI}_{N,e} +  \sigma^{SI}_{N,o} \,,
\end{align}
where the vector SI cross-section is given by (see~\cite{Lewin:1995rx, Yaguna:2015mva})
\begin{equation}
\label{eq:SI-vector}
\sigma^{SI}_{N,o} = \dfrac{G_F^2\, m_N^2}{4\,\pi A^2}(\cos\theta_L^2+\cos\theta_R^2)^2
\left[(1-4\sin^2\theta_W)Z-(A-Z)\right]^2 \,,
\end{equation}
and the scalar SI cross-section is given by
\begin{equation}
\label{eq:SI-scalar}
\sigma^{SI}_{N,e} \approx \dfrac{ m_N^4 f_N^2}{\pi v^2}
\left(\dfrac{g_{\chi_1^0\chi_1^0 h_1}}{m_{h_1}^2} + \dfrac{g_{\chi_1^0\chi_1^0 h_2}}{m_{h_2}^2}\right)^2\,,
\end{equation}
with DM coupling to the Higgs fields written as
\begin{equation}
g_{\chi_1^0\chi_1^0 h_k} = \dfrac{-i}{\sqrt{2}}\sin\theta_R\left(h_d\cos\theta_L Z^H_{k1} + h_c\sin\theta_L Z^H_{k2}\right)\,.
\end{equation}
In the model presented in Ref.~\cite{Yaguna:2015mva}, which is a limiting case of our model and where the DM particle is mainly singlet, direct detection bounds imply that the mixing angles $\theta_{L,R}$ need to be very small. In that case, the only way to achieve the current value of the relic density of DM is via coannihilations, forcing the neutral fermions to be quasi-degenerate, $M_{\Psi} \sim M_N$. In this work, however,  that is not the case because of the presence of the new scalar $S$, which facilitates the depletion of DM during early stages of the universe. This allows us to obtain the correct relic density without coannihilations playing an important role, as we will show numerically in the next section. 

Finally, the axial-vector interaction of DM with nucleons yields the SD scattering cross-section, which has been probed by several experiments such as XENON1T~\cite{Aprile:2019dbj} and LUX~\cite{Akerib:2017kat}. As we will see in Sec.~\ref{sec:scan}, the SD interactions provide less stringent restrictions on our scenario than the SI interactions.

\subsection{Numerical results}
\label{sec:scan}

\begin{figure}[h]
\centering
\includegraphics[scale=0.43]{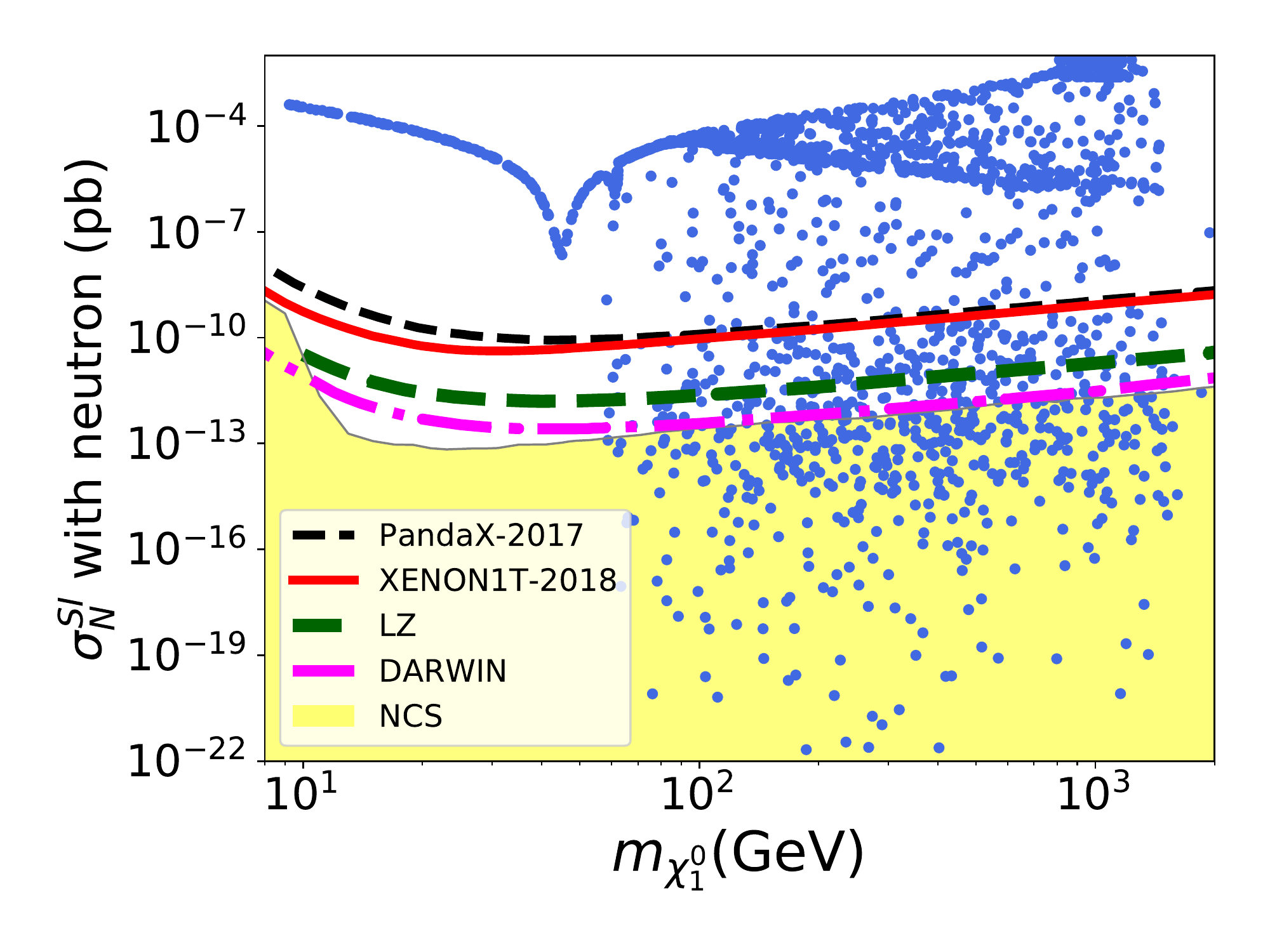}
\includegraphics[scale=0.43]{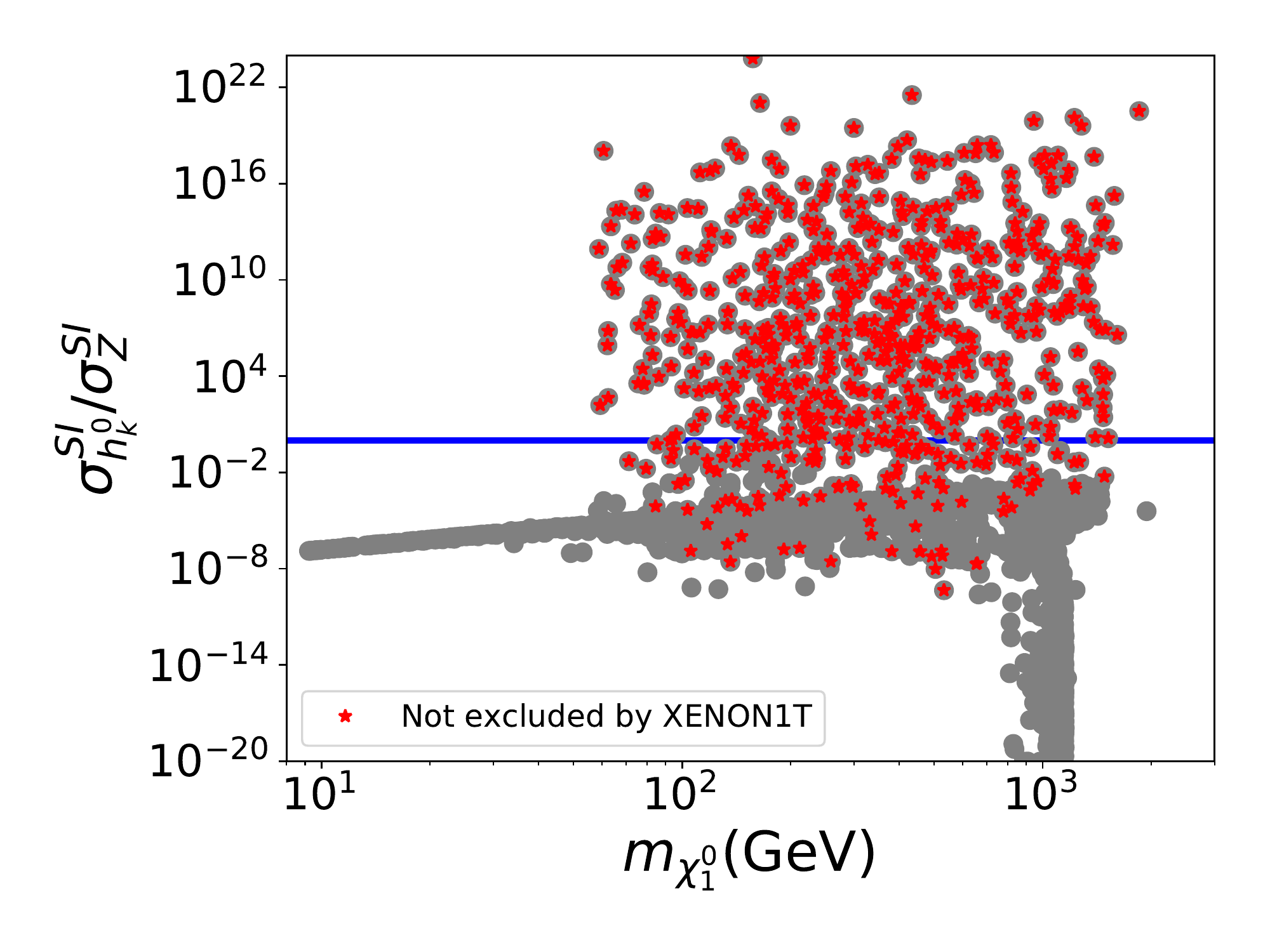}
\caption{Left: The SI cross-section (blue dots) and the current experimental constraints from XENON1T~\cite{Aprile:2018dbl}, PandaX~\cite{Cui:2017nnn}, and prospects from LZ~\cite{Akerib:2018lyp} and DARWIN~\cite{Aalbers:2016jon}. 
We also show the Neutrino Coherent Scattering (NCS)~\cite{Cushman:2013zza, Billard:2013qya}. Right: The grey dots show the ratio between the scalar and vector SI cross-sections. The red stars are those models that are below the XENON1T limit}.
\label{fig:SI}
\end{figure}

In order to study the phenomenology of this model, we have performed a random scan of the parameter space, varying the free parameters as described in Table~\ref{tab:scan}.
\begin{table}[t]
\centering
\begin{tabular}{|c|c|} 
\hline
Parameter & Range\\
\hline
$M_\Psi$ (GeV) & $10^{2}-10^{4}$ \\
$m_{\sigma_i}$ (GeV) & $10^{3}-2\times 10^{4}$ \\
$v_S$ (GeV) & $10^{2}-10^{5}$ \\
$|h_c|$, $|h_d|$ & $10^{-6}-3$ \\
$\lambda^{H\sigma_i}$, $\lambda^S$, $\lambda^{S\sigma_i}$, $\lambda_{SH}$, $\lambda^{\sigma_i}$ & $10^{-4}-3$ \\
$|h_a^{2i,3i}|$ & $10^{-6}-1$\\
\hline
\end{tabular}
\label{tab:scan}
\caption{Scan range of the free parameters of our model. The remaining parameters are obtained from the ones in this table. In particular, $h_b^{\alpha 1}$ and $h_b^{\alpha 2}$ are fixed by Eq.~\eqref{eq:ha-and-hb}, resulting in the range $10^{-8}< h_b^{\alpha 1,\alpha 2}<1$.}
\end{table}
We implemented the model in~\texttt{SARAH}~\cite{Staub:2008uz,Staub:2009bi,Staub:2010jh,Staub:2012pb,Staub:2013tta}, coupled to the \texttt{SPheno}~\cite{Porod:2003um,Porod:2011nf} routines. In order to obtain the DM relic density, we used~\texttt{MicrOMEGAs 4.2.5}~\cite{Belanger:2006is}, which takes into account all the possible channels contributing to the relic density, mentioned in Sec.~\ref{sec:h2-relic-density}, including special processes such as coannihilations and resonances~\cite{Griest:1990kh}. We selected the models that fulfill the current value $\Omega_\chi h^2 = (0.120 \pm 0.001)\; \text{to}\; 3\sigma$~\cite{Aghanim:2018eyx} and, at the same time, reproduce the neutrino parameters described in Sec.~\ref{sec:neutrinos}.
For those points, we computed the SI DM-nucleus scattering cross-section, shown in Eq.~\eqref{eq:full-SI}, and checked it against the current experimental bounds of XENON1T~\cite{Aprile:2018dbl}, PandaX~\cite{Cui:2017nnn}, and prospect bounds of LZ~\cite{Akerib:2018lyp} and DARWIN~\cite{Aalbers:2016jon}. The results are shown on the left panel of Fig.~\ref{fig:SI}.  
We analyzed the vector and scalar SI cross-sections separately in order to discern the behavior of these two contributions to the total SI cross-section and we found that the vector contribution dominates the region above the XENON1T limit. Therefore, it needs to be suppressed in order to escape the current bounds. The majority of models with large vector SI cross-section are excluded; these correspond to large mixing angles $\theta_{L,\,R}$, as seen from Eq.~\eqref{eq:SI-vector}. Thus, the viable DM candidate needs to be mostly singlet in order to suppress the $Z$-portal and fulfill the current direct detection constraints. An analytic estimate tells us that this is achieved by requiring $\cos\theta_{L,\,R} \leqslant 0.1$.
For illustrative purposes, on the right panel of Fig.~\ref{fig:SI}, we show the ratio between the scalar and vector SI cross-sections. The red stars correspond to the viable models that are not excluded by XENON1T. These models have a sizable scalar contribution ($\sigma^{\text{SI}}_{N,e}=\sigma^{\text{SI}}_{h_k^0}$) and low vector cross-section ($\sigma^{\text{SI}}_{N,o}=\sigma^{\text{SI}}_{Z}$), except for some points that fall below the blue line which have a dominant vector cross-section while escaping the DD bounds as analized in Ref.~\cite{Yaguna:2015mva}. 

\begin{figure}
\centering
\includegraphics[scale=0.43]{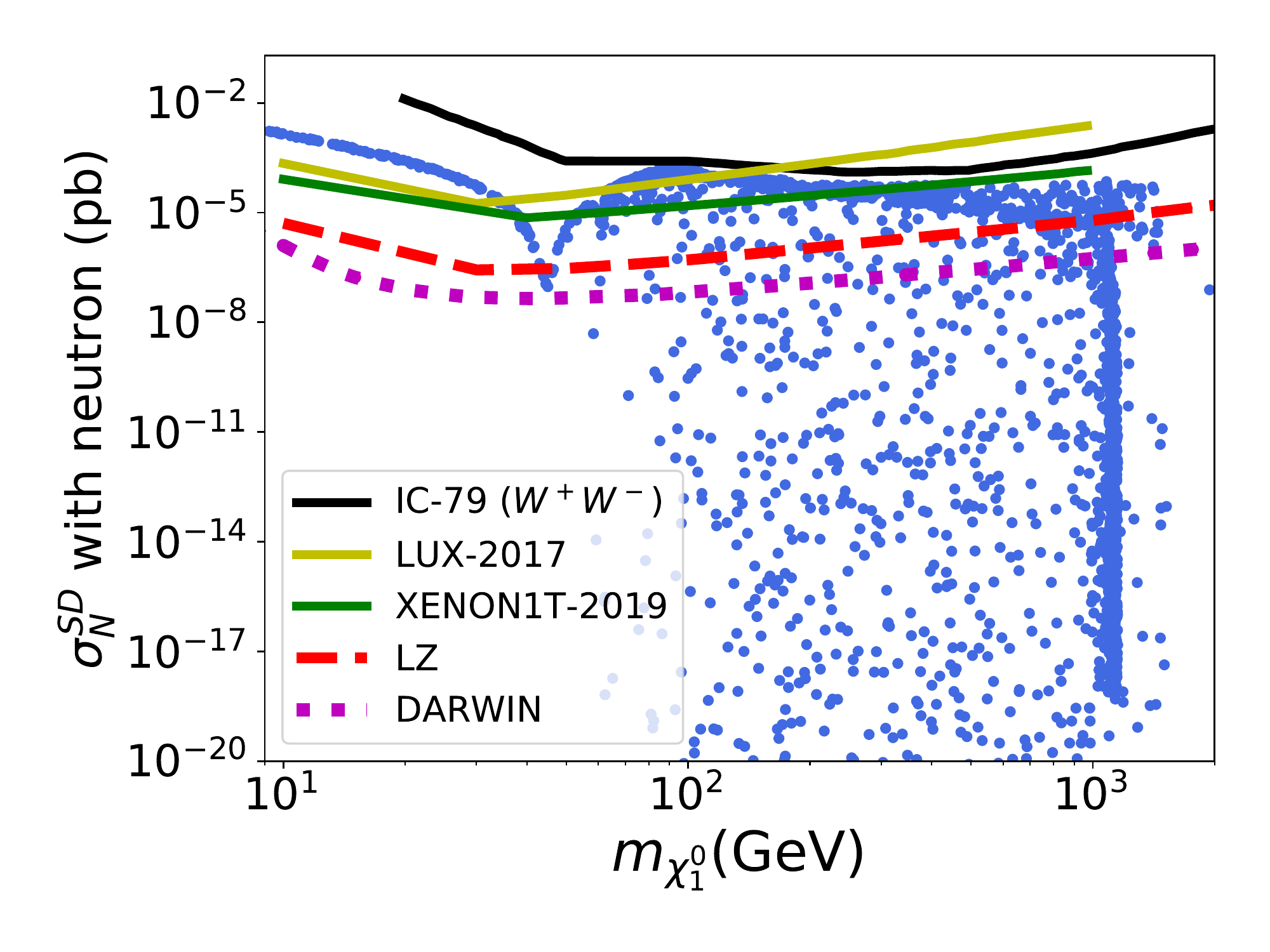}
\includegraphics[scale=0.43]{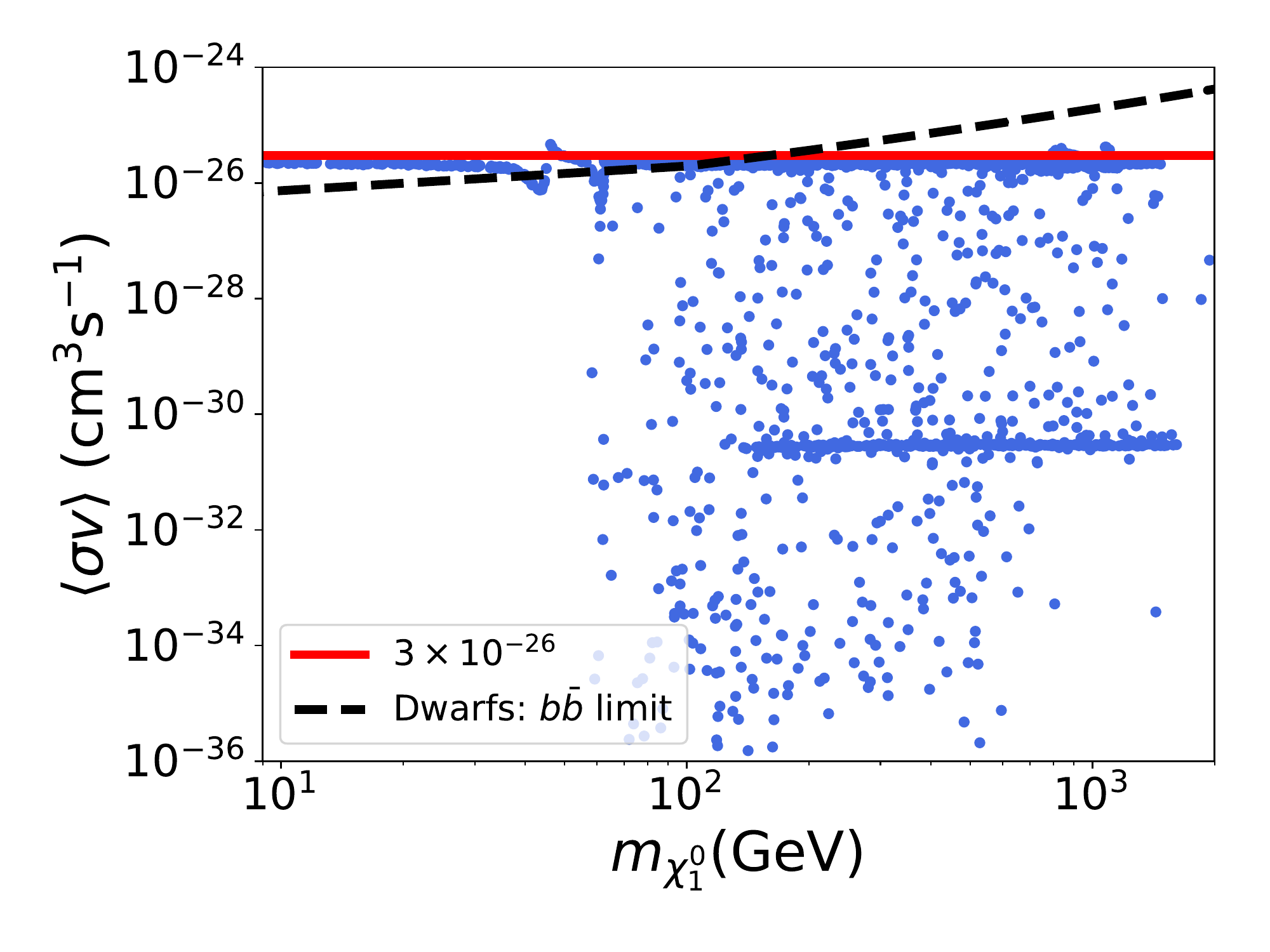}
\caption{Left: WIMP-neutron SD cross-sections and the current experimental constraints from XENON1T~\cite{Aprile:2019dbj}, LUX~\cite{Akerib:2017kat}, Ice-cube~\cite{2013PhRvL.110m1302A}, and prospects as LZ~\cite{Akerib:2018lyp} and DARWIN~\cite{Aalbers:2016jon}. Right: Annihilation cross-section today. We also show the typical thermal value $\langle\sigma v\rangle\sim 3\times 10^{-26}\, \text{cm}^3\,\text{s}^{-1}$   in the early universe and the experimental limit for DM annihilation into $b\bar{b}$ in dwarf galaxies (dSphs)~\cite{Ackermann:2015zua}.}
\label{fig:SD-sv}
\end{figure}

In order to complete this analysis, we show on the left panel of Fig.~\ref{fig:SD-sv} the behavior of the WIMP-neutron spin-dependent (SD) cross-section for the points in the parameter space that yield the expected value of the relic abundance and reproduce the neutrino physics. 
We also show the IceCube~\cite{2013PhRvL.110m1302A} limits on the $W^+W^{-}$ channel (black solid
line) for DM annihilation at the sun, the limits from LUX~\cite{Akerib:2017kat} (yellow solid line), the current and most restricted limits from XENON1T~\cite{Aprile:2019dbj} (green solid line), and the expected sensitivities of LZ~\cite{Akerib:2018lyp} (red dashed line) and DARWIN~\cite{Aalbers:2016jon} (magenta dot line). As in the case of the SI cross-section, we can see that DARWIN~\cite{Aalbers:2016jon} could probe some region of the parameter space of this model. Evidently, the points that are below the neutrino floor could be confused with the neutrino scattering with nucleons and they would need a special analysis that is beyond the scope of this work.

Finally, on the right panel of Fig.~\ref{fig:SD-sv}, we show today's annihilation cross-section times velocity, $\sigma v$, which allows us to look at indirect detection (ID) constraints. We used ~\texttt{MicrOMEGAs 4.2.5} to compute $\sigma v$ today for each point of the scan. Notice that these results show the expected suppression due to the $p-$wave nature of the DM annihilation. 
Therefore, the  indirect DM detection prospects of this model are significantly low. For instance, the points with $m_{\chi_1^0}\lesssim 100$ GeV could have a large branching ratio of the annihilation channel $\chi_1^0\bar{\chi_1}^0\to b\bar{b}$, leading to DM annihilation into $b\bar{b}$ signals from dwarf galaxies (dSphs)~\cite{Ackermann:2015zua}. However, as seen previously, those points are already excluded by DD. 
Combining the direct and indirect detection constraints, we conclude that all models with $m_{\chi^0_1} \lesssim 65$ GeV are excluded, except for the funnel region due to resonances with the $Z$ and the $h_1$ gauge bosons.

\begin{figure}
\centering
\includegraphics[scale=0.5]{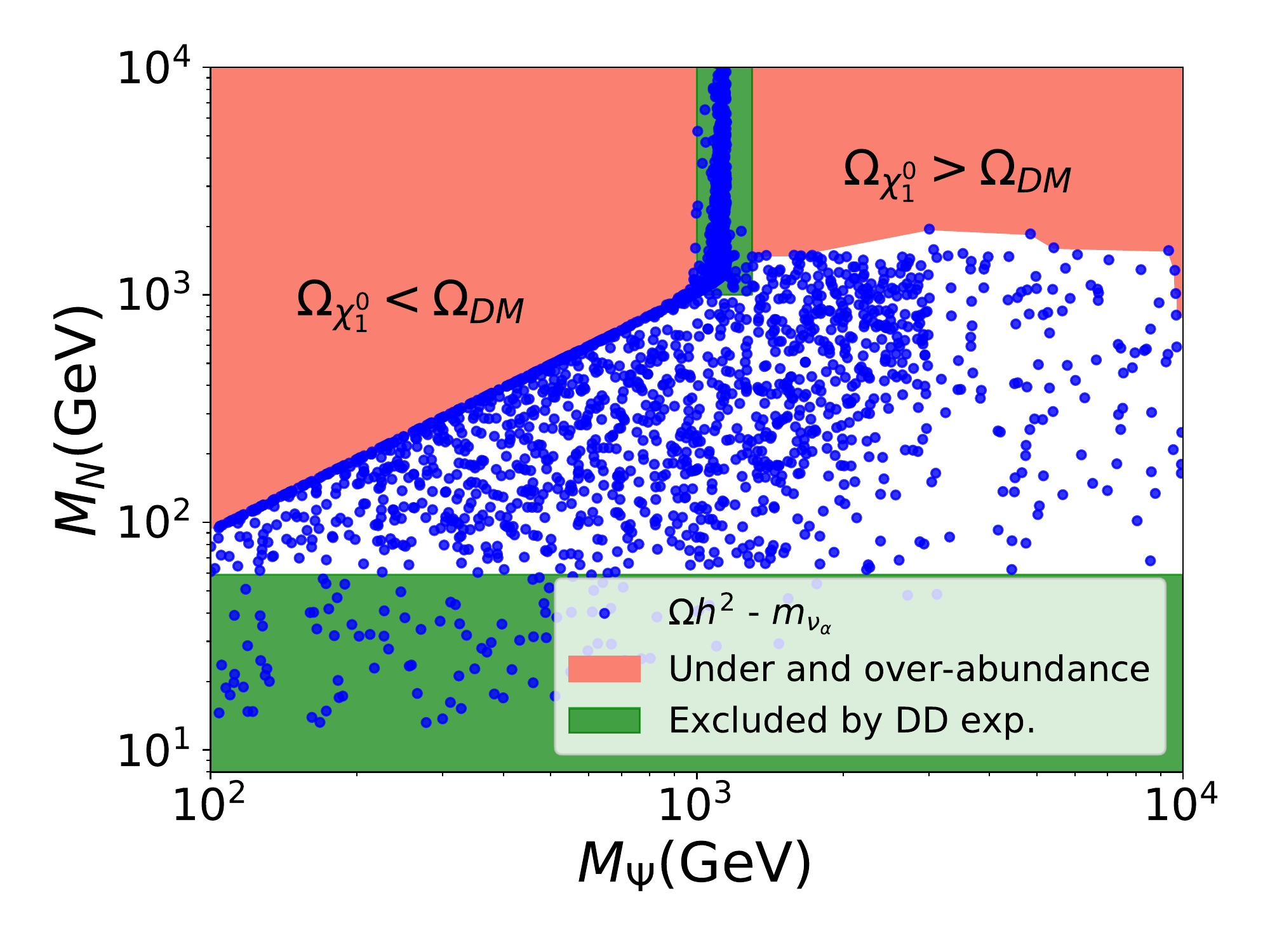}
\caption{$M_{\Psi} - M_N$ plane for the scan done in this work. The blue dots give the correct relic abundance and reproduce the neutrino parameters. The pink-shaded region corresponds to overabundance or underabundance of DM. The green-shaded region is excluded by direct detection (DD) experiments.}
\label{fig:MSvsMD}
\end{figure}

Following the analysis described above, we project the scanned points on the $M_\Psi - M_N$  plane and show it in Fig.~\ref{fig:MSvsMD}. In the figure, the blue dots show the models that yield the correct value of the relic density and reproduce the neutrino parameters while the green-shaded region is excluded by DD experiments. The pink shade shows the region where a larger, $\Omega_{\chi_1^0}>\Omega_{\text{DM}}$, or or smaller, $\Omega_{\chi_1^0}<\Omega_\text{DM}$, relic density is obtained.
Notice that, in our scenario, the region that leads to the correct abundance is much wider than in the Majorana fermion case~\cite{Cohen:2011ec,Cheung:2013dua,Calibbi:2015nha,Restrepo:2015ura} and the original SD$^3$M proposal~\cite{Yaguna:2015mva}, allowing the parameter $M_N$ up to $2$~TeV as shown in the figure.

In general, for models with $M_\Psi > M_N$,  outside the region where coannihilations are relevant, the relic density is set through different channels in the early universe.   
As argued in Sec.~\ref{sec:h2-relic-density}, the main process is $\chi_1^0\bar{\chi_1}^0\to h_2h_2$. In that case, we have checked that the expression shown in Eq.~\eqref{eq:RelicAbundance} is in good agreement.
Finally, in the coannihilation region  ($M_\Psi\approx M_N$), the main contributions to the relic density come from $\chi_2^0\chi^+ \to f\bar{f'} (\nu\bar{e}, u\bar{d},...)$ mediated by the $W^{\pm}$ boson, followed by $\chi_2^0\bar{\chi_2}^0 \to f\bar{f}$ and $\chi^-\chi^+ \to f\bar{f}$. In this limit, processes involving the DM particle have a negligible contribution to the relic density because they are characterized by low Yukawa couplings as described in Ref.~\cite{Yaguna:2015mva}. 

Finally, regarding collider searches, this scenario can be tested using the search for electroweak production of charginos $\chi^{\pm}$ decaying in final states with two leptons and missing transverse momentum in $pp$ collisions at the LHC~\cite{ATLAS:2019cfv}. 
Those analyses have been done in the context of simplified SUSY models and can be recast in this analysis. 
The observed limit rules out masses up to $120$ GeV for $\chi_1^0$, with  $m_{\chi^{\pm}}$ $\lesssim 420$ GeV. 
However, in that case, the $\chi^{\pm}$ are wino-like particles with a production cross-section that is larger than in this model, where $\chi^{\pm}$ are higgsino-like particles ($SU(2)$ doublet). With this in mind, we estimate that the low production rate decreases the values of $M_N$ that can be probed to $M_N \lesssim 100$~GeV, which makes it inapplicable to our allowed region of parameter space. Nevertheless, a better analysis needs to be done in this direction and we leave it for future work.

\section{Conclusions}

After several decades of model building and experimental search, the nature of DM is still unknown. Among the many possible scenarios, a Dirac fermion is a viable candidate within the singlet-doublet scenario SD${}^3$M~\cite{Yaguna:2015mva}. 
In this paper, we have minimally extended that model in order to generate Dirac neutrino masses via the radiative seesaw mechanism. 
We have scanned the parameter space requiring that the correct DM relic abundance and current neutrino data are reproduced while being compatible with direct detection experiments. 
We found a DM candidate that is a Dirac fermion resulting from a mixture of new singlet-doublet fields with mass  $65\,\text{GeV}\lesssim m_{\chi_1^0}\lesssim 1.1\,\text{TeV}$.
The inclusion of the new scalar $S$ opens a new portal, which, in association with the vector $Z$ portal, contributes to the $\text{SI}$ cross-section, widening the allowed parameter space while opening up the testing prospects in future direct detection experiments. 
Additionally, unlike in the original SD${}^3$M proposal, coannilitations do not play a central role in setting the relic abundance in our model.
Regarding indirect detection, this framework does not provide clear prospective signatures since the annihilation cross-section is $p$-wave suppressed.

\section{Acknowledgments}

We are grateful to Oscar Zapata and Federico Von der Pahlen for enlightening discussions. DR is partially  supported by COLCIENCIAS grant 111577657253 and Sostenibilidad-UdeA.
AR is supported by COLCIENCIAS through the ESTANCIAS POSTDOCTORALES program 2017, and by the APS  International Research Travel Award Program.
The research of WT is supported by the College of Arts and Sciences of Loyola University Chicago. 
AR is grateful to LUC for the hospitality during the final stage of this work.

\appendix

\section{Lagrangian in terms of Weyl spinors }
\label{sec:l-weyl-spinors}
The Lagrangian in Eq.~\eqref{eq:full-lagrangian-Dirac} can be written in terms of chiral spinors as follows:
\begin{align}
\label{eq:mass-term}
\mathcal{L} \supset &
-M_{\Psi}\, \overline{\Psi}\Psi
= -M_{\Psi}\,(\overline{\Psi^0},\overline{\Psi^-})\begin{pmatrix}
\Psi^0 \\ \Psi^-
\end{pmatrix}\nonumber\\
&=-M_{\Psi}\,(\overline{\Psi^0}\Psi^0 + \overline{\Psi^-}\Psi^-)
= -M_{\Psi}\,((\Psi^0)^\dagger \gamma_0 \Psi^0 + (\Psi^-)^\dagger \gamma_0 \Psi^-)\nonumber\\
&=-M_{\Psi}\,\left[(\Psi^0_R,\Psi^0_L )^\dagger \begin{pmatrix}
0 & 1\\
1 & 0
\end{pmatrix}\begin{pmatrix}
\Psi^0_R \\ \Psi^0_L
\end{pmatrix}
+(\Psi^-_R,\Psi^-_L )^\dagger \begin{pmatrix}
0 & 1 \\
1 & 0
\end{pmatrix} 
\begin{pmatrix}
\Psi^-_R \\ \Psi^-_L
\end{pmatrix}\right]\nonumber\\
&= -M_{\Psi}\,\left[(\Psi^0_L )^\dagger\Psi^0_R + (\Psi^-_L )^\dagger\Psi^-_R + \text{h.c.}\right]\nonumber\\
&=-M_{\Psi}\,\left[(-(\Psi^-_R)^\dagger,(\Psi^0_R)^\dagger) \begin{pmatrix}
0 & -1\\
1 & 0
\end{pmatrix} 
\begin{pmatrix}
\Psi^0_L\\ \Psi^-_L
\end{pmatrix} + \text{h.c.} \right]
=-M_{\Psi}\, \left[\widetilde{(\Psi_R)} \cdot \Psi_L + \text{h.c.}\right]\,,
\end{align}
where, the dot product represents the $i\sigma_2$ matrix and $\widetilde{(\Psi_R)}=(-(\Psi^-_R)^\dagger,(\Psi^0_R)^\dagger)^T$ , $\Psi_L=(\Psi^0_L,\Psi^-_L)^T$ are two chiral doublets of $SU(2)$ with opposite hypercharge.
In the same way, 
\begin{align}
\label{eq:hd-term}
\mathcal{L} & \supset \, 
h_d\,\overline{\Psi}\widetilde{H}\psi_{R} 
+ \text{h.c.} 
= h_d\, (\overline{\Psi^0},\overline{\Psi^-})\begin{pmatrix}
(H^0)^* \\ -H^-
\end{pmatrix}\psi_R + \text{h.c.} 
=h_d\, \left[(\Psi^0_L)^\dagger(H^0)^*\psi_R  - (\Psi^-_L)^\dagger H^-\psi_R  + \text{h.c.}\right] \nonumber \\
&=h_d\, \left[(\psi_R )^\dagger H^0 \Psi_L^0 - (\psi_R )^\dagger H^+ \Psi_L^-  + \text{h.c.}\right]
=h_d\,(\psi_R )^\dagger (H^+,H^0)\begin{pmatrix}
0 & -1 \\
1 & 0
\end{pmatrix}\begin{pmatrix}
\Psi_L^0 \\ \Psi_L^-
\end{pmatrix} + \text{h.c.} \nonumber \\
&=h_d\, (\psi_R )^\dagger H\cdot\Psi_L + \text{h.c.} 
\end{align}
\begin{align}
\label{eq:ha-term}
\mathcal{L} & \supset\,
 h_a^{\beta i}\, \overline{L}_\beta \Psi \sigma_i + \text{h.c.} 
=  h_a^{\beta i}\,\left((\nu_L)_\beta^\dagger\Psi_R^0 + (e_L)_\beta^{\dagger}\Psi_R^- \right)\sigma_i  + \text{h.c.}
=  h_a^{\beta i}\,\left((\Psi_R^0)^\dagger(\nu_L)_\beta + (\Psi_R^-)^{\dagger}(e_L)_\beta \right)\sigma_i  + \text{h.c.}\nonumber\\
&= h_a^{\beta i}\,(-(\Psi_R^-)^{\dagger},(\Psi_R^0)^\dagger)\begin{pmatrix}
0 & -1 \\
1 & 0
\end{pmatrix}\begin{pmatrix}
(\nu_L)_\beta \\ (e_L)_\beta
\end{pmatrix}\sigma_i + \text{h.c.} 
=  h_a^{\beta i}\,\widetilde{(\Psi_R)}\cdot L_\beta \sigma_i + \text{h.c.}
\end{align}
\begin{align}
\label{eq:hbc-term}
\mathcal{L}  & \supset\, 
 h_b^{\alpha i}\, \overline{\psi_L}\,\nu_{R\alpha} \sigma_i +
 h_c\,\overline{\psi_R}\,\psi_L S  + \text{h.c.}
 = h_b^{\alpha i}\, \overline{\psi}\,P_R \nu_{\alpha} \sigma_i +
 h_c\,\overline{\psi}\,P_L\psi S  + \text{h.c.} \nonumber\\
 & = h_b^{\alpha i}\,(\psi_R , \psi_L)^{\dagger}\begin{pmatrix}
 0&1\\
 1&0
 \end{pmatrix}\begin{pmatrix}
 \nu_{R\alpha} \\ 0
 \end{pmatrix}\sigma_i
 + h_c\,(\psi_R , \psi_L)^{\dagger}\begin{pmatrix}
 0&1\\
 1&0
 \end{pmatrix}\begin{pmatrix}
 0 \\ \psi_L
 \end{pmatrix}S + \text{h.c.} \nonumber \\
 & = h_b^{\alpha i} ({\psi}_{L})^\dagger \nu_{R\alpha} \sigma_i
+ h_c\, ({\psi}_{R})^\dagger\psi_{L} S + \text{h.c.}
\end{align}
Therefore, replacing the Eqs.~\eqref{eq:mass-term},\eqref{eq:hd-term},~\eqref{eq:ha-term}, and~\eqref{eq:hbc-term} in Eq.~\eqref{eq:full-lagrangian-Dirac}, we obtain
\begin{align}
\mathcal{L} \supset &
-M_{\Psi}\, \left[\widetilde{(\Psi_R)} \cdot \Psi_L +\text{h.c.}\right]  -V(H,\sigma_i, S)\nonumber  \\
&+  \left[ h_a^{\beta i}\,\widetilde{(\Psi_R)}\cdot L_\beta \sigma_i
+ h_b^{\alpha i} ({\psi}_{L})^\dagger \nu_{R\alpha} \sigma_i
+ h_c\, ({\psi}_{R})^\dagger\psi_{L} S 
+ h_d\, (\psi_R )^\dagger H\cdot\Psi_L 
+ \text{h.c.}\right] \,.
\end{align}

\bibliographystyle{jhep}
\bibliography{references}

\end{document}